\documentclass[12pt]{iopart}
\pdfoutput=1 
\usepackage{color}
\usepackage{graphicx}
\usepackage{iopams}

\begin{document}

\title[Efficiency of transport in periodic potentials]{Efficiency of transport in periodic potentials:  dichotomous noise contra deterministic force}
\author{J. Spiechowicz,  J. {\L}uczka and L. Machura}
\address{Institute of Physics, University of Silesia, 40-007  Katowice, Poland}
\address{Silesian Center for Education and Interdisciplinary Research,\\ University of Silesia, 41-500 Chorz{\'o}w, Poland}

\ead{lukasz.machura@us.edu.pl}

\begin{abstract}

We study transport of an inertial Brownian particle moving in a \emph{symmetric} and periodic one-dimensional potential,  and subjected to both a \emph{symmetric}, unbiased external harmonic force as well as  biased dichotomic noise $\eta(t)$ also known as a random telegraph signal or a two state continuous-time Markov process. In doing so, we concentrate on the previously reported regime [J. Spiechowicz \textit{et al.}, Phys. Rev. E \textbf{90}, 032104 (2014)] for which non-negative biased noise $\eta(t)$ in the form of generalized white Poissonian noise can induce anomalous transport processes similar to those generated by a deterministic constant force $F=\langle \eta(t) \rangle $ but significantly more effective than $F$, i.e. the particle moves much faster, the velocity fluctuations are noticeable reduced and the transport efficiency is enhanced several times. Here, we confirm this result for the case of dichotomous fluctuations which in contrast to white Poissonian noise can assume positive as well as negative values and examine the role of thermal noise in the observed phenomenon. We focus our attention on the impact of bidirectionality of dichotomous fluctuations and reveal that  the effect of non-equilibrium noise enhanced efficiency is still detectable. This result may explain transport phenomena occurring in strongly fluctuating environments of both a  physical and biological origin. Our predictions can be corroborated experimentally by use of a set-up that consists of a resistively and capacitively shunted Josephson junction.
\end{abstract}


\maketitle

\section{Introduction}
\begin{figure}[b]
	\centering
	\includegraphics[width=0.45\linewidth]{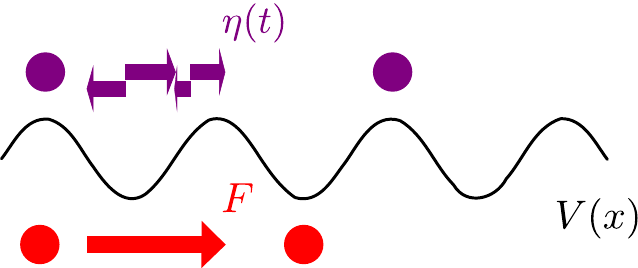}
	\caption{Brownian particle moving in 
	symmetric and periodic potential $V(x)$, diven by a harmonic force $A\cos{(\omega t)}$ 
	and subjected to a deterministic constant force $F$ can be transported in much more
	effective way when $F$ is replaced by dichotomous noise $\eta(t)$ of equal average bias
	$\langle \eta(t) \rangle = F$.}
	\label{moti}
\end{figure}
The question whether deterministic forces are preferred for transporting particles or maybe random perturbations should be applied is not simple to answer in a unique way. The first alternative has been commonly used because it is deterministic and therefore predictable. The latter seems to be bizarrely risky because it is random  and therefore unpredictable.  This opinion is based on our everyday experience  that randomness, stochasticity  and noise are  uncontrolled and therefore can lead to unintended consequences. However, there are phenomena in Nature where randomness  plays a constructive role. Examples are: chemical reaction driven by thermal fluctuations \cite{reactions}, stochastic resonance \cite{stochres} or Brownian motors \cite{LucTal2000,ratchets}. In the paper \cite{spiechPRE}, influence of two forces on transport of the Brownian particle (motor) have been compared with respect to its effectiveness. One perturbation is a deterministic static force $F>0$ and the other is Poisson noise $\eta(t) \ge 0$ of the same mean value as $F$, i.e. $\langle \eta(t) \rangle =F$. It has been shown that there are parameter regimes in which system driven by noise $\eta(t)$ responses more effectively than a system driven by a static force $F$ in the sense that the stationary average velocity of the Brownian motor is several times greater, its fluctuations are reduced distinctly and the transport efficiency becomes greatly enhanced. In this paper, we analyse a similar problem by replacing the Poisson process with the dichotomous process. It is an extension of the previous studies in two aspects. Realizations of Poisson noise considered in Ref. \cite{spiechPRE} consist of only positive $\delta$-kicks of an infinite amplitude which act on the system in an infinitesimally short period of time \cite{spiechJSTATMECH, spiechSCRIPTA}. On the other hand, dichotomous noise can assume both positive and negative values with random non-zero residence times. Moreover, in some limiting cases, dichotomous noise tends to either Gaussian white noise or Poisson white noise with positive as well as with both positive-and-negative $\delta$-kicks. In this way, we can study a wider class of random perturbations and check to what extend the phenomenon  of the noise enhanced effectiveness is universal and robust. Because the dichotomous process is characterised by four parameters (two possible states and two mean waiting times in these states), we expect to detect distinctly novel transport behaviour. Exploiting advanced numerical simulations with CUDA environment implemented on a modern desktop GPU \cite{spiechCPC}, we search the full parameter space of the system and demonstrate how transport of the Brownian motor can be controlled by dichotomous noise parameters and reveal regimes in which the motor efficiency is strongly enhanced.

The remaining part of the paper is organised in the following way. In Section 2 we present details of the model in terms of the Langevin equation for the Brownian motor. The next Section provides a description of dichotomous noise together with its typical realizations. Section 4 is devoted to analysis of main transport characteristics, namely the long-time averaged velocity of the motor, its fluctuations and efficiency. Conclusions contained in Section 5 summarise the work.
\section{Driven noisy dynamics}
The archetype model of transport of the  Brownian particle of mass $M$ moving in a periodic potential $V(x)=V(x+L)$ of period $L$ and driven by both an  external time-periodic force $G(t)=G(t+\cal{T})$ of period $\cal{T}$  and a static force $F$ reads
\begin{equation}
\label{eq:Langevin}
M \ddot{x} + \Gamma \dot{x} = - V'(x) + G(t) + \sqrt{2\Gamma k_B T} \xi(t) + F, 
\end{equation}
where the dot and the prime denotes the differentiation with respect to time $t$ and the particle position $x\equiv x(t)$, respectively. The parameter $\Gamma$ is the friction coefficient and $k_B$ is the Boltzmann constant. Contact with thermostat of temperature $T$ is described by thermal fluctuations modelled here by Gaussian white noise $\xi(t)$ of zero mean and unit intensity, i.e.
\begin{equation}
\langle \xi(t) \rangle = 0, \quad \langle \xi(t)\xi(s) \rangle = \delta(t-s).
\end{equation}
The r.h.s. of (\ref{eq:Langevin}) describes the influence of  various forces on the dynamics. All except one are unbiased and their mean values are zero: $\langle V'(x) \rangle = 0$ over its period $L$, $\langle G(t)\rangle = 0$ over its period $\cal{T}$ and symmetric thermal noise $\langle \xi(t)\rangle = 0$ over its realizations. The only biased force is the static perturbation $F$. In the special case when both $V(x)$ and $G(t)$ are symmetric and additionally $F\equiv 0$, there is no directed transport of the particle in the asymptotic long time limit. When the space reflection symmetry of $V(x)$ and/or time reflection symmetry of $G(t)$ is broken transport can be induced even if $F = 0$ \cite{spiechPRB}. In this work we assume the symmetric potential $V(x)$ and the symmetric  driving $G(t)$. Therefore to induce a directed motion of the motor, we have to postulate that the static force $F \ne 0$. 

The symmetric potential $V(x)$ is assumed to be in the simple form
\begin{equation}	
	V(x) =  V_0 \sin(2\pi x/L) 
\end{equation}	
and the time-periodic force $G(t)$ is chosen to be
\begin{equation}
\label{eq:driving}	
	G(t) = A \cos(\Omega t).  
\end{equation}	
The model (\ref{eq:Langevin})-(\ref{eq:driving}) has been intensively studied in the literature \cite{MacKos2007,KosMac2008,acta_jj}. Here we replace the deterministic static force $F$ by its random counterpart  $\eta(t)$ and compare their impact on effectiveness of the particle transport. As a stochastic force $\eta(t)$ we 
consider a two state continuous-time Markov process, namely, dichotomous noise also known as a random telegraph signal. 
To be precise we study the following Langevin equation  
\begin{equation}
\label{eq:Langevin2}
M \ddot{x} + \Gamma \dot{x} = - V'(x) + G(t) + \sqrt{2\Gamma k_B T} \xi(t) + \eta(t)
\end{equation}
with the constraint on the mean value of dichotomous noise $\langle \eta(t) \rangle = F$.
\section{Dichotomous noise}	
Dichotomous noise \cite{BroHan1984,KulCze1996,bena} assumes two states
\begin{equation}
\label{eq:dichnoise}
\eta(t) = \{b_-, b_+\}, \quad b_+>b_-.
\end{equation}
The states  $b_-$ and $b_+$ are specified by any real numbers with the above restriction $b_+>b_-$. In a typical scenario $b_-<0$ and $b_+>0$. The probabilities of transition per unit time from one state to the other are given by the relations
\begin{eqnarray}
\label{eq:dnmuab}
Pr(b_- \to b_+) &=& \mu_- = \frac{1}{\tau_-},\nonumber\\
Pr(b_+ \to b_-) &=& \mu_+ = \frac{1}{\tau_+}
\end{eqnarray}
where in turn $\tau_-$ and $\tau_+$ are mean waiting times in the states $b_-$ and $b_+$, respectively. The mean value and the autocorrelation function of noise (\ref{eq:dichnoise}) read
\begin{eqnarray}
\label{eq:dnmoments}
\langle \eta(t) \rangle = \frac{b_+ \mu_- + b_- \mu_+}{\mu_+ + \mu_-} = \frac{b_+ \tau_{+} + b_- \tau_{-}}{\tau_{+} + \tau_{-}}, \\
\langle \eta(t) \eta(s) \rangle = \frac{Q}{\tau} e^{-|t-s|/\tau}, 
\end{eqnarray}
where the intensity $Q$ and the correlation time $\tau$ are expressed by the relations
\begin{equation}
\label{eq:dnintensity}
Q = \mu_+ \mu_- \tau^3 (b_+ -b_-)^2 , \quad \frac{1}{\tau} = \frac{1}{\tau_+} + \frac{1}{\tau_-}. 
\end{equation}
\begin{figure}[t]
	\centering
	\includegraphics[width=0.45\linewidth]{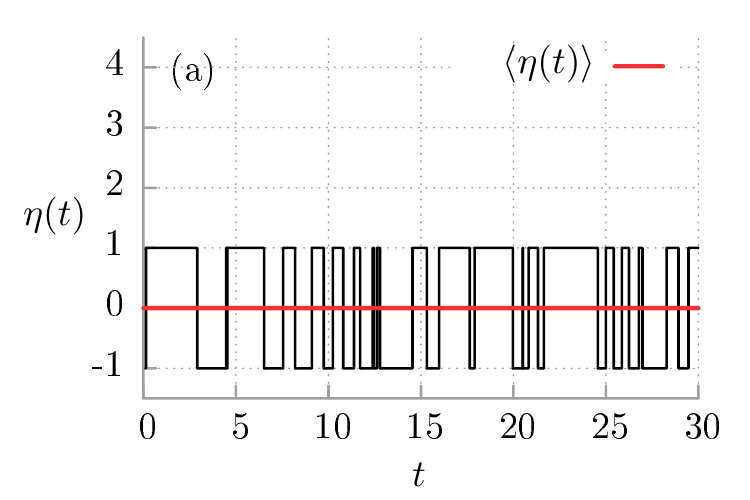} \\
	\includegraphics[width=0.45\linewidth]{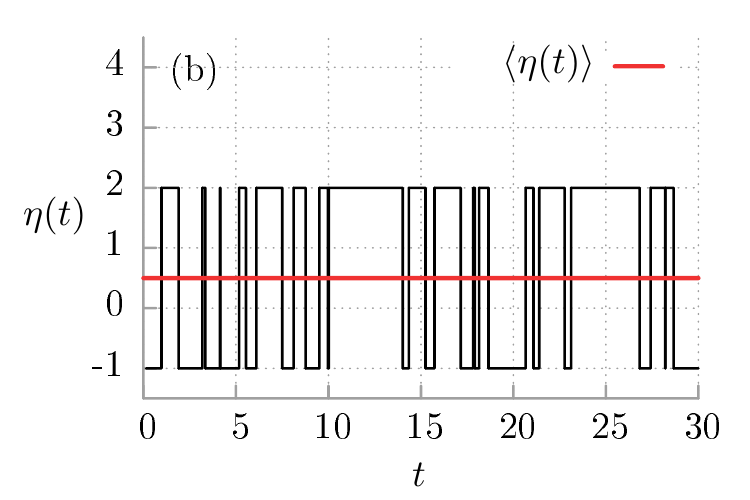}
	\includegraphics[width=0.45\linewidth]{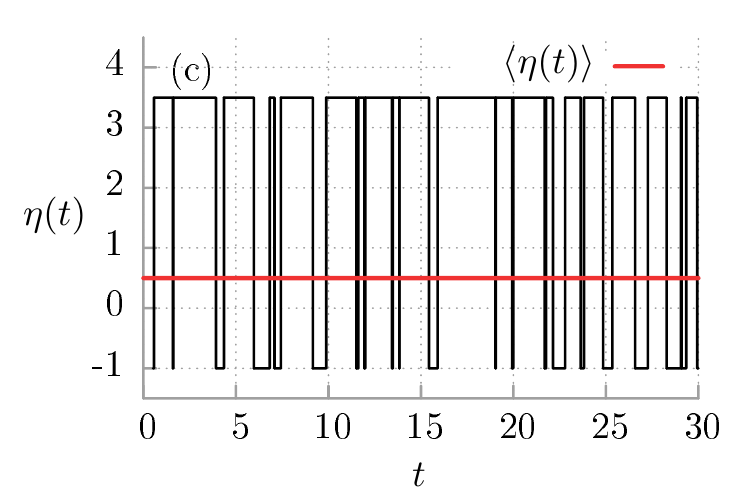}
	\caption{Three illustrative realizations of  dichotomous noise also known as the two-state continuous-time Markov process or a random telegraph signal. Panel (a): the symmetric dichotomous fluctuations with $b_+ = 1$, $b_- = -1$, $\mu_+ = 1$ and $\mu_- = 1$. Panel (b) and (c) depict the asymmetric process with $b_- = -1$ and $\mu_- = 1$. The remaining parameters in plot (b) and (c) are $b_+ = 2$, $\mu_+ = 1$ and $b_+ = 3.5$, $\mu_+ = 2$, respectively.}
	\label{fig1}
\end{figure} 

In figure \ref{fig1} we depict three illustrative realizations of dichotomous noise. Panel (a) presents symmetric fluctuations of the vanishing average value. Plots (b) and (c) show the asymmetric dichotomous process with the fixed mean value $\langle \eta(t) \rangle = 0.5 > 0$. The reader may observe there the impact of changes in the parameters characterizing the noise, the states $b_+$ and $b_-$ as well as the mean transition probabilities $\mu_+$ and $\mu_-$ on its overall realizations. 

In appropriate limits dichotomous noise tends to Poisson or Gaussian white noise \cite{broeck}. In the the first case $b_+ \to \infty$ and $\tau_+ \to 0$ in such a way that $b_+\tau_+=const.$ On the other hand, the Gaussian white 
noise is achieved when $b_+ \to \infty, b_- \to -\infty, \tau_+ \to 0, \tau_- \to 0$ with $b_+b_-\tau=const$.
\subsection{Dimensionless Langevin equations}
In the following we make use of the dimensionless notation introduced in \cite{MacKos2004}. Therefore time will be scaled with the characteristic time $\tau_0^2 = M L^2 / V_0$ and the x--coordinate of the Brownian particle with the characteristic length $L$, i.e. $\hat{t} = t/\tau_0$ and $\hat{x} = x/L$. With these assumptions (\ref{eq:Langevin}) and (\ref{eq:Langevin2}) can be converted into its dimensionless form. The corresponding dimensionless Langevin equations read
\begin{eqnarray}
\ddot{\hat{x}}(\hat{t}) + \gamma \dot{\hat{x}}(\hat{t}) =-  \hat{V}'(\hat{x}) + a \cos (\omega \hat{t}\, ) + \sqrt{2\gamma D_T} \, {\hat\xi}({\hat t}\,) + f, \label{eq:dLangevin1}\\
\ddot{\hat{x}}(\hat{t}) + \gamma \dot{\hat{x}}(\hat{t}) =-  \hat{V}'(\hat{x}) +  a \cos (\omega \hat{t}\, ) + \sqrt{2\gamma D_T} \,{\hat\xi}({\hat t}\,) + \hat{\eta}({\hat t}\,). \label{eq:dLangevin2}
\end{eqnarray}
In this scaling the particle mass $m =1$  and the dimensionless  
friction coefficient \mbox{$\gamma = \tau_0 \Gamma / M$}. The rescaled potential $\hat{V}(\hat{x}) = V(L\hat{x})/V_0 = \sin(2\pi \hat{x})$ is characterized by the unit period:  $\hat{V}(\hat{x}) = \hat{V}(\hat{x} +1)$ and barrier height $V_0 = 2$. Other re-scaled dimensionless parameters are: the amplitude $a = LA/V_0$ and the angular frequency $\omega = \tau_0 \Omega$ of the time-periodic driving. We introduced the dimensionless thermal noise intensity \mbox{$D_T = k_BT/V_0$}, so that  Gaussian white noise $\hat{\xi}(\hat{t})$ of vanishing mean $\langle\hat{\xi}(\hat{t})\rangle =0$ possesses the auto-correlation function $\langle \hat{\xi}(\hat{t})\hat{\xi}(\hat{s}) \rangle = \delta(\hat{t} - \hat{s})$. The rescaled static force is $f = F L / V_0$. The dimensionless dichotomous noise now takes values $\hat{\eta}(\hat{t}) = \{b_- L / V_0, b_+ L / V_0\} \equiv \{ \hat{b}_-, \hat{b}_+\}$. Its mean value is $\langle \hat{\eta} (\hat{t}) \rangle = (L/V_0) \langle \eta(t) \rangle$ and the correlation function $\langle \hat{\eta}(\hat{t}) \hat{\eta}(\hat{s}) \rangle = (\hat{Q} / \hat{\tau}) \exp[-|\hat{t} - \hat{s}|/\hat{\tau}]$, with the intensity $\hat{Q} = \hat{\mu}_+ \hat{\mu}_- \hat{\tau}^3 (\hat{b}_+ - \hat{b}_-)^2 $, where $\hat{\mu}_+ =\tau_0 \mu_+, \hat{\mu}_- =\tau_0 \mu_-$ and the correlation time $\hat{\tau} = \tau / \tau_0$. From now on for the sake of simplicity we shall skip all the hats in the above equations and parameters.

The significance of the investigated model is due to its widespread representation in experimentally accessible physical systems which can be described by use of the above equations. Among others, prominent examples that come to mind are the following: pendulums \cite{gitterman2010}, super-ionic conductors \cite{fulde1977}, charge density waves \cite{gruner1981}, Josephson junctions \cite{kautz1996}, Frenkel-Kontorova lattices \cite{braun1998}, ad-atoms on solid surfaces \cite{guantes2001} and cold atoms in optical lattices \cite{renzon1,renzon2,denisov2014}.

\section{Transport characteristics}
\label{sec4}
The most essential quantifier for characterization of transport processes occurring in periodic systems described by the driven noisy dynamics (\ref{eq:dLangevin1}) or (\ref{eq:dLangevin2}) is an asymptotic long time \emph{average velocity} $\langle v \rangle$ given by the following formula
\begin{equation}
	\langle v \rangle = \lim_{t\to\infty} \frac{\omega}{2\pi} \int_{t}^{t+2\pi/\omega} 
{\mathbb E}[v(s)] \; ds, 
\end{equation}
where ${\mathbb E}[v(s)]$ stands for the ensemble average over realizations of random forces and noises as well as over the set of initial conditions. The latter is mandatory especially in the deterministic limit of vanishing thermal and dichotomous noise since then the system is typically \emph{non-ergodic}. An additional averaging procedure over the period $2\pi/\omega$ of the external harmonic driving is necessary to extract only the time-independent component of the Brownian particle velocity which in the asymptotic long time limit assumes its periodicity \cite{acta_jj,jung1993,spiechNJP}.

Apart from this very basic transport characteristic indicating its directed nature there are other characteristics which are useful to describe its effectiveness \cite{jung1996,linke2005,MacKos2004}. Among them particularly 
important are \emph{fluctuations of velocity} estimated by the variance 
\begin{equation}
	\sigma_v^2 = \langle v^2 \rangle - \langle v \rangle^2.
\end{equation}
Since typically in the long time limit the instantaneous Brownian particle velocity $v(t) \in [\langle v \rangle - \sigma_v, \langle v \rangle + \sigma_v]$ we easily see that when the velocity fluctuations are sufficiently large, 
$\sigma_v > |\langle v \rangle|$, then for a certain period of time the particle may move in the direction opposite to its average velocity, the transport is intermittent and therefore not optimal. An ideal situation occurs when the particle travels with the high speed $|\langle v \rangle|$ and simultaneously the fluctuations of velocity $\sigma_v$ are small. Then the transport process is systematic and efficient.

Finally, in order to measure the effectiveness of transport we utilize the so called \emph{Stokes efficiency} \cite{MacKos2004,linke2005,spiechNJP, oster} which 
is evaluated as the ratio of the dissipated power $P_{out} = f_v\langle v \rangle$ associated with the directional movement against the mean viscous force $f_v = \gamma \langle v \rangle$ to the input power $P_{in}$ \cite{MacKos2004}
\begin{equation}
	\varepsilon_S = \frac{P_{out}}{P_{in}} = \frac{\langle v \rangle^2}{\langle v \rangle^2 + \sigma_v^2 - D_T} = \frac{\langle v \rangle^2}{\langle v^2 \rangle - D_T}.
\end{equation}
Here, the form of the input power $P_{in}=\langle [a \cos(\omega t) +f]v\rangle$ and $P_{in}=\langle [a \cos(\omega t) +\eta(t)]v\rangle$ supplied to the system by the external 
forces follows from the energy balance of the underlying equations of motion (\ref{eq:dLangevin1}) and (\ref{eq:dLangevin2}). 
We note that this definition of the efficiency agrees well with our previous statement: the transport is optimized in the regimes which \emph{maximize} the directed velocity and \emph{minimize} its fluctuations. 

\subsection{General remarks on dynamics and transport properties}
The deterministic system corresponding to (\ref{eq:dLangevin1})  
has a three-dimensional phase space, namely, $\{x, \dot x, \omega t\}$. It is the minimal phase space dimension necessary for it to display chaotic evolution which is an important feature for anomalous transport to occur. Its dynamics is able to exhibit a diversity of behavior in phase space as a function of the system parameters. Trajectories can be periodic, quasiperiodic and chaotic. Typically, there are two possible dynamical states of the system: a locked state, in which the particle oscillates mostly within one or several potential wells and a running one.  Moreover, one can distinguish two classes of running states: either the  particle moves forward without any back-turns or it undergoes frequent oscillations and back-scattering events. The running states are crucial for the directed transport properties. 

In general, the force-velocity curve $\langle v \rangle = \langle v \rangle(f)$  is a non-linear function of the constant force $f$. From the symmetries of the underlying Langevin equation of motion (\ref{eq:dLangevin1}) it follows that it is odd in the force  $f$, i.e. $\langle v \rangle(-f)=-\langle v \rangle(f)$ and as a consequence $\langle v \rangle(f=0)=0$. So, we need $f \ne 0$ to break the symmetry of the system and to induce directed transport in the asymptotic long-time regime. Typically, the velocity is an increasing function of the force $f$. Such regimes correspond to a normal transport behaviour. More interesting are, however, regimes of anomalous transport, exhibiting an absolute negative mobility (ANM) when $\langle v\rangle <0$ for $f>0$. In accordance with the Le Chatelier-Braun principle \cite{landau} it is already known that the key ingredient for the occurrence of ANM is that the system is driven far away from thermal equilibrium into a time-dependent non-equilibrium state in such a way that it is exhibiting a vanishing, unbiased non-equilibrium response. This goal may be realized for example by applying the unbiased time periodic force $G(t)$ as we did it in our work.

The underlying deterministic dynamics can be chaotic and
therefore a fractal structures of certain domains must be expected to exist. The richness and diversity of subtle areas where ANM can be detected is large. The regions of ANM form stripes, fibres and islands \cite{KosMac2008,ANMcolor,janus}.  At 'low' temperature, we observe the refined structure with many narrow, slim and twisted regions of this effect. The occurrence of ANM may be governed by two different mechanisms. In some regimes it is solely induced by thermal equilibrium fluctuations, i.e. the effect is absent for vanishing thermal fluctuations $D_T=0$. This situation is nevertheless rooted in
the complex deterministic structure of the non-linear dynamics
governed by a variety stable and unstable orbits \cite{MacKos2007}. In other regimes, anomalous transport may also occur in the noiseless, deterministic system and can be understood if one studies the structure of the existing attractors and the corresponding basins of attraction \cite{acta_jj}. In this case, if temperature is increased, this subtle structure of ANM is increasingly smeared out and becomes smoother. Many previously existing domains of ANM start to shrink or vanish altogether. We detect some few robust regimes for which anomalous transport persists. Outside these domains, a normal response to the load $f$ is found and it dominates in the parameter space.

\subsection{Details of simulations}
Unluckily, the Fokker-Planck equation corresponding to the driven Langevin dynamics described by (\ref{eq:dLangevin1}) and (\ref{eq:dLangevin2}) cannot be handled by any known analytical methods. For this reason, in order to study transport properties of the system,  we have performed comprehensive numerical simulations of the investigated models. In particular, we have integrated the Langevin equations (\ref{eq:dLangevin1}) and (\ref{eq:dLangevin2}) by employing a weak version of the stochastic second order predictor corrector algorithm with a time step typically set to about $10^{-3} \cdot 2\pi/\omega$. We have chosen the initial coordinates $x(0)$ and velocities $v(0)$ equally distributed over the intervals $[0,1]$ and $[-2,2]$, respectively. Quantities of interest were ensemble averaged over $10^3 - 10^4$ different trajectories which evolved over $10^3 - 10^4$ periods of the external harmonic driving. All calculations have been done with the aid of a CUDA environment implemented on a modern desktop GPU. This procedure allowed for a speed-up of a factor of the order $10^{3}$ times as compared to a common present-day CPU method \cite{spiechCPC}.

\subsection{Results}
Despite the use of innovative computer technologies the system described by (\ref{eq:dLangevin2}) has a 9-dimensional parameter space $\{\gamma, a, \omega, D_T, f, b_+, b_-, \mu_+, \mu_-\}$ which unfortunately is still too large to analyse numerically in a systematic fashion. We have found that the normal transport regime (i.e. $\langle v\rangle >0$ for $f>0$)  dominates in the parameter space. However, we can also identify a remarkable and distinct property of anomalous transport, namely, ANM. In both regimes, one may find parameter regions where the impact of dichotomous noise $\eta(t)$ is destructive, i.e. the absolute value of the directed velocity $\langle v\rangle$ is suppressed in comparison to the deterministic force $f = \langle \eta(t)\rangle$. However, there are also areas where $\eta(t)$ acts constructively. Because normal transport regimes are not so interesting as ANM ones, below we will analyse only the last scenarios. Moreover, we restrict our attention to a regime of the non-equilibrium noise enhanced ANM phenomenon studied in detail in Ref. \cite{spiechPRE}. Unless stated otherwise, this case corresponds to the following set of parameters $\{\gamma, a, \omega, D_T\} = \{1.546, 8.95, 3.77, 0.001\}$. This regime seems to be optimal in the sense that the negative mobility is most profound in a relatively large domain with relatively large values of the dimensionless velocity. On the one hand, it allows to compare the influence of Poisson and dichotomous noise. Moreover, it gives us possibility to state to what extend this effect is universal with respect to different kind of stochastic forcing.

\begin{figure}[t]
	\centering
	\includegraphics[width=0.45\linewidth]{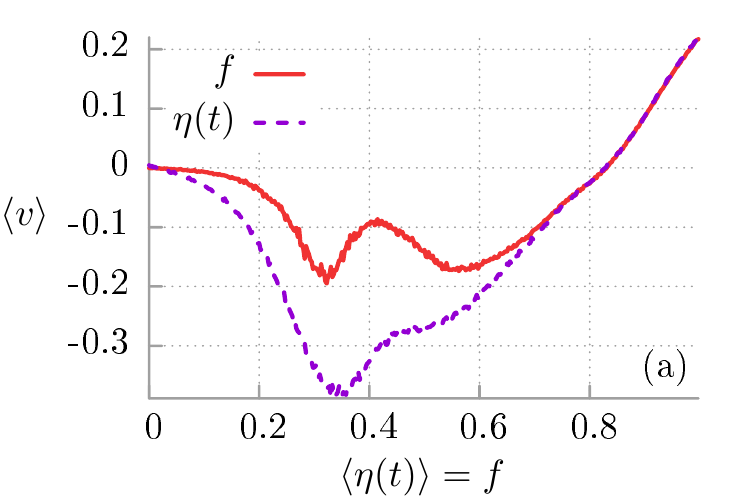}
	\includegraphics[width=0.45\linewidth]{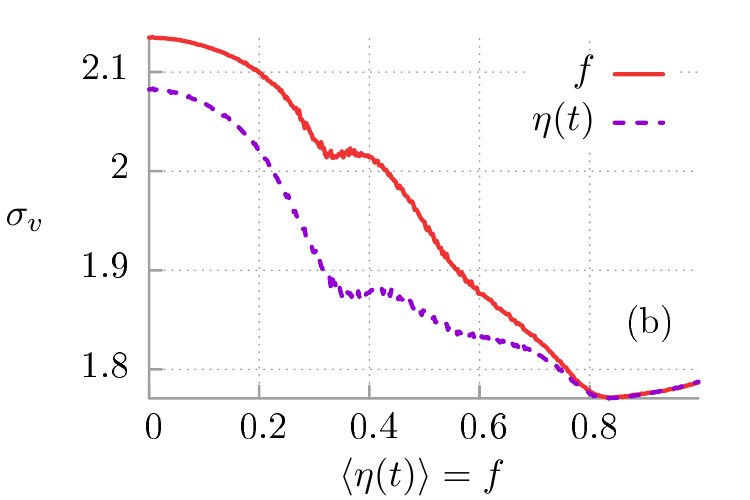}\\
	\includegraphics[width=0.45\linewidth]{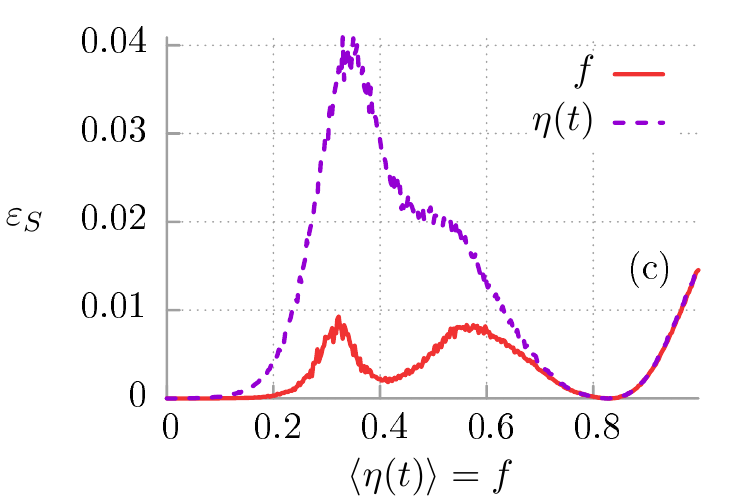}
	\caption{Efficiency of transport in the absolute negative mobility regime versus the biasing force of either a deterministic or a stochastic origin $\langle \eta(t) \rangle = f > 0$. The random perturbation is modelled by dichotomous noise. Panel (a): the asymptotic long time average velocity $\langle v \rangle$. Panel (b): the velocity fluctuations $\sigma_v$. Panel (c): the Stokes efficiency $\varepsilon_S$. Parameters are $\gamma = 1.546$, $a = 8.95$, $\omega = 3.77$, $D_T = 0.001$, $b_+ = 0.85$, $\mu_+ = 8.7$, $\mu_- = 0.708$. The other dichotomous state $b_-$ is changing so that the condition $\langle \eta(t) \rangle = f$ is satisfied.}
	\label{fig2}
\end{figure}
\begin{figure}[t]
	\centering
	\includegraphics[width=0.45\linewidth]{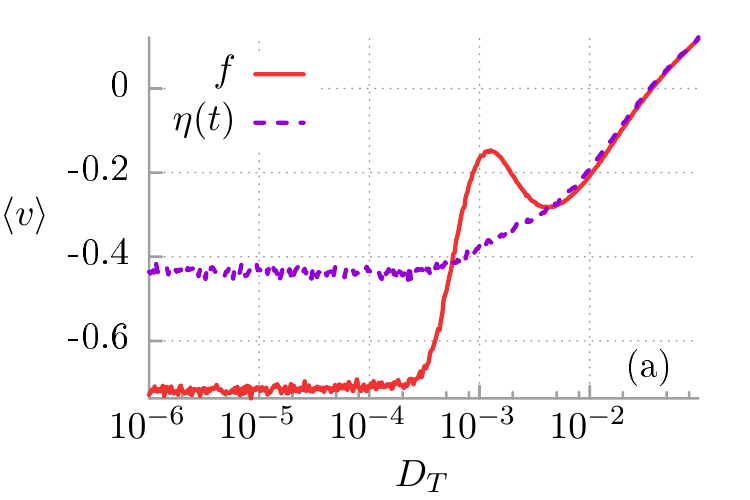}
	\includegraphics[width=0.45\linewidth]{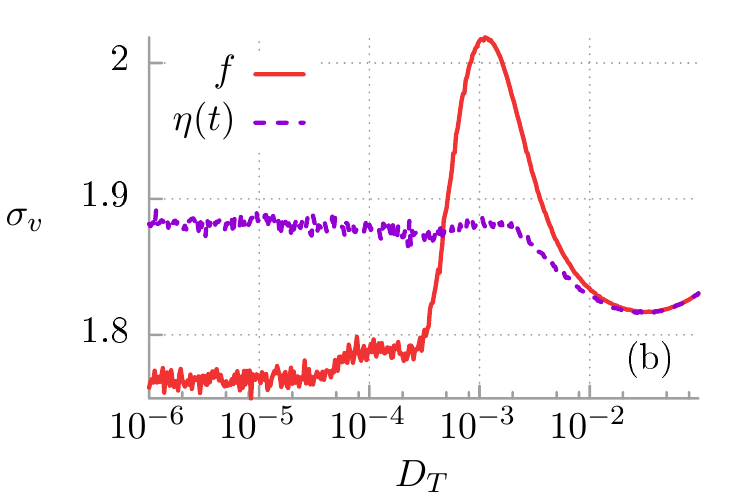}\\
	\includegraphics[width=0.45\linewidth]{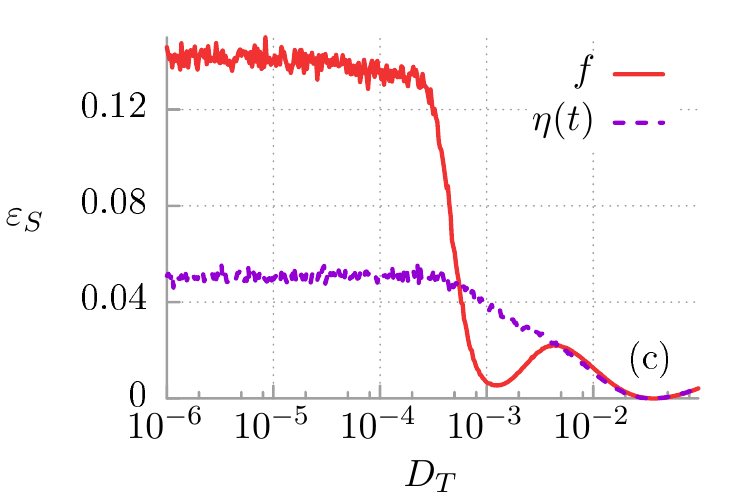}
	\caption{Impact of temperature $D_T \propto T$ on the transport efficiency in the absolute negative mobility regime. The biasing force is fixed to $\langle \eta(t) \rangle = f = 0.34$. Parameters modelling dichotomous fluctuations are as follows $b_+ = 0.85$, $b_- = 0.3$, $\mu_+ = 8.7$, $\mu_- = 0.708$, for others see figure \ref{fig2}.}
	\label{fig3}
\end{figure}
We start with the asymptotic long time ensemble averaged velocity $\langle v \rangle$. In particular, we investigate whether the transport velocity can be enhanced when the deterministic force $f$ acting on the Brownian particle is replaced by the stochastic perturbation $\eta(t)$ in the form of dichotomous noise. This is indeed the case as we show it in figure \ref{fig2}(a). The average velocity is depicted as a function of the biasing force of either deterministic or stochastic origin $\langle \eta(t) \rangle = f > 0$. Since in the vicinity of vanishing bias this transport characteristic points into the negative direction the presented regime corresponds to the absolute negative mobility phenomenon \cite{MacKos2007}. 
Moreover, when the Brownian particle is subjected to  dichotomous noise instead of the constant static force the effect can be observed for slightly higher values of the bias. 
Finally, in both cases there exists an optimal value for the bias $\langle \eta(t) \rangle = f = 0.34$ at which the average velocity takes its minimal value. Surely the most interesting observation is that for dichotomous noise this minimum is 
nearly two times more pronounced than for the corresponding deterministic force.

In panels \ref{fig2}(b) and \ref{fig2}(c) we present the measures characterizing quality of the Brownian particle transport versus the biasing force $f$ or $\langle \eta(t)\rangle$. In particular, plot \ref{fig2}(b) depicts velocity fluctuations $\sigma_v$. Unfortunately,  
they are an order of magnitude greater than the average velocity $\langle v \rangle$ presented in figure \ref{fig2}(a). However, still for the case of dichotomous noise $\eta(t)$ 
the velocity fluctuations are significantly smaller than when the Brownian particle is subjected to the constant static force $f$. Notably, the counter-intuitive effect of \emph{reduction of fluctuations by fluctuating perturbation} $\eta(t)$ is observed for a wide interval of the stochastic driving. This fact, together with the enlargement of the negative response  illustrated in panel \ref{fig2}(a), is reflected in 
the dependence of the Stokes efficiency $\varepsilon_S$ on the biasing mechanism which is shown in figure \ref{fig2}(c). In both cases there exists an optimal value of efficiency.   However, in the case of dichotomous noise $\eta(t)$ this quantifier grows by a factor four over the value obtained by application of the static force $f$.
\begin{figure}[t]
	\centering
	\includegraphics[width=0.6\linewidth]{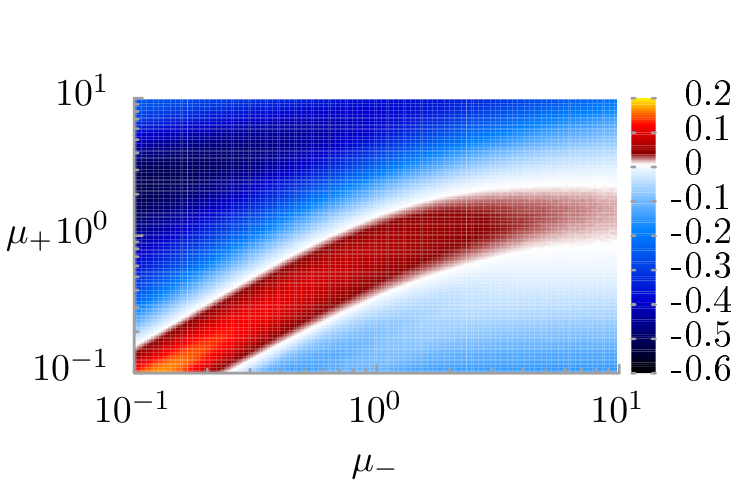}
	\caption{The asymptotic long time average velocity $\langle v \rangle$ of the driven Brownian particle subjected to dichotomous noise $\eta(t)$ presented as a function of the transition probabilities $\mu_+$ and $\mu_-$. The mean value of the stochastic perturbation is fixed to $\langle \eta(t) \rangle = 0.34$ and $b_+ = 0.85$. Other parameters of the model are the same as in figure \ref{fig2}.}
	\label{fig4}
\end{figure}

We now turn to the analysis of the impact of  temperature $D_T \propto T$ on the transport efficiency in the remarkable regime of the absolute negative mobility phenomenon. In figure \ref{fig3}(a) we present the dependence of the 
average velocity $\langle v \rangle$ on the thermal noise intensity $D_T$. Since in the limiting case of the very low temperatures the particle response measured as its velocity is still negative we conclude that this anomalous transport effect has its origin in the complex, deterministic, albeit 
chaotic dynamics \cite{MacKos2007,speer2007}. It is significant that for  low temperatures the particle velocity is larger in the case of the constant, static force $f$. However, evidently there is a finite window of the thermal noise intensities $D_T$ for which the particle moves faster when it is subjected 
to dichotomous fluctuations $\eta(t)$. For the opposite limiting case of high temperature there is no distinction between the response induced by these two types of forces. In panel 3(b) we depict the velocity fluctuations $\sigma_v$ 
versus thermal noise intensity $D_T$. Similarly as in the previous case, for low temperatures fluctuations of velocity are smaller when the particle is propelled by the constant bias $f$. However, as thermal noise intensity grows we observe also a rapid increase of the velocity fluctuations up to 
their local maximum which is reached for $D_T \approx 0.001$. On the other hand, when dichotomous noise acts on the particle then  fluctuations of velocity are almost constant in the entire presented interval of temperature. In particular, for the critical value $D_T \approx 0.001$ they are noticeable lower. This fact has further consequences on the transport efficiency $\varepsilon_S$ which is plotted versus thermal noise intensity $D_T$ in panel \ref{fig3}(c). One can note that generally the constant bias $f$ induces more effective transport than dichotomous noise $\eta(t)$. Nonetheless, due to the explained fine tuning between equilibrium (thermal) and non-equilibrium (dichotomous) fluctuations the opposite scenario can also be observed. This remark lead us to the novel conclusion that the phenomenon of a non-equilibrium 
noise enhanced transport efficiency is \emph{assisted} by thermal fluctuations.

\begin{figure}[t]
	\centering
	\includegraphics[width=0.45\linewidth]{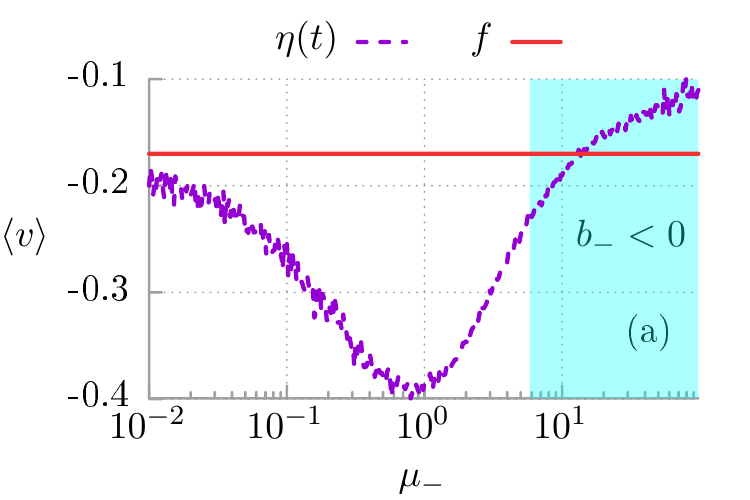}
	\includegraphics[width=0.45\linewidth]{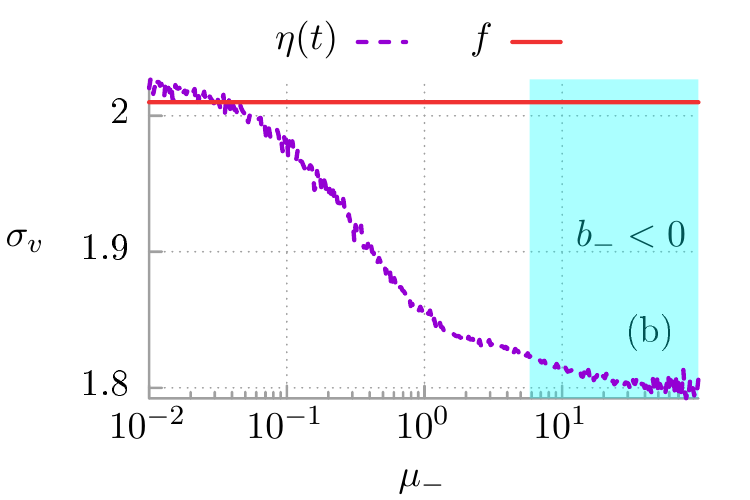}\\
	\includegraphics[width=0.45\linewidth]{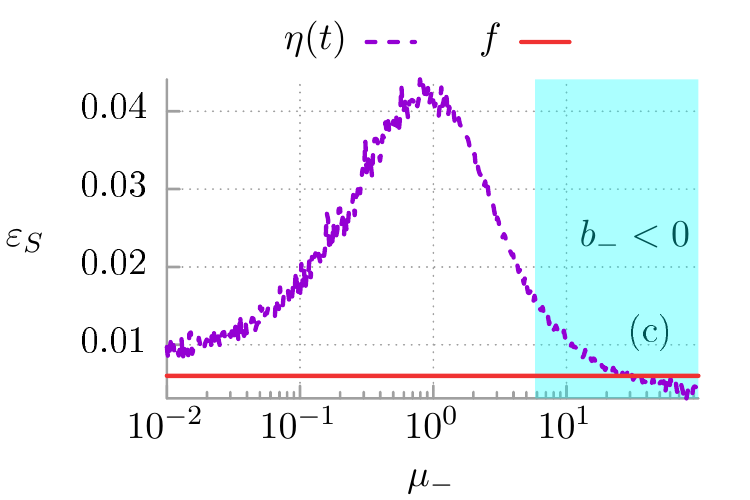}
	\caption{Efficiency of transport in the absolute negative mobility regime versus the transition probability $\mu_-$. Panel (a): the asymptotic long time average velocity $\langle v \rangle$. Panel (b): the velocity fluctuations $\sigma_v$. Panel (c): the Stokes efficiency $\varepsilon_S$. The biasing force is fixed to $\langle \eta(t) \rangle = f = 0.34$. The other parameters characterizing dichotomous noise are to $b_+ = 0.85$ and $\mu_+ = 8.7$. For the rest read figure \ref{fig2}. The region corresponding to the bidirectional dichotomous noise ($b_+ > 0$ and $b_- < 0$) is indicated with the light blue colour.}
	\label{fig5}
\end{figure}
Since dichotomous noise can assume both positive and negative values, the question whether it is possible to manipulate the direction of the Brownian particle transport naturally arises. We answer this one in positive in figure \ref{fig4} where we plot the asymptotic long time average velocity $\langle v \rangle$ as a function of both the transition probabilities 
$\mu_+$ and $\mu_-$ for the fixed mean value of the dichotomous noise $\langle \eta(t) \rangle = 0.34$ and $b_+ = 0.85$. Regions of the positive and negative velocity are clearly observed. Moreover, surprisingly, for the presented parameter regime the latter are more common. When the transition 
probabilities are comparable then the transport occurs in a normal regime. Conversely, when there is a large difference between them then we deal with the absolute negative mobility response. The phenomenon of multiple velocity reversals 
which is illustrated here is known for the driven periodic systems \cite{spiechPRB, bartussek1994, jung1996, mateos2000}. By the fine tuning of the parameters describing  dichotomous noise one can control the direction of transport in this set-up.

The phenomenon of a non-equilibrium noise enhanced efficiency of Brownian motors operating in the micro-scale domain has been recently illustrated for the case of unidirectional fluctuations modelled by non-negative white Poissonan noise 
\cite{spiechPRE}. We now ask whether such a phenomenon can be observed when the Brownian particle is subjected to bidirectional random perturbation which can assume both positive as well as negative values. Such a case is extremely important since it may help to elevate the understanding of the transport properties not only in physical but especially in biological systems, where instead of systematic, deterministic load there are random forces usually acting without any specific direction \cite{bressloff2013}. The studied dichotomous noise is one of the simplest models which takes into account these prominent aspects of dynamics. In 
figure \ref{fig5} we show that, indeed, for selected parameter regimes the phenomenon of noise enhanced efficiency of the motor can be observed even if amplitudes of dichotomous noise are of opposite signs. The region corresponding to  the $\eta(t)$-amplitudes of opposite signs is indicated there with the light blue colour. Although the enhancement of transport efficiency is peculiarly noticeable when both of the 
dichotomous states are positive this effect still survives in the case of the bidirectional perturbations. In the presented regime the optimal situation for this phenomenon to occur corresponds to the two positive states $b_+ > 0$ and $b_- > 0$. 
However, we remind that we restricted our analysis to only one set of parameters of the multidimensional space and therefore almost certainly there are regimes for which the reversed scenario is observed. This results shed new light 
on the possible role of fluctuations and random perturbations in transport phenomena occurring in the nano and micro-scale.
\section{Summary}
\label{sec5}
We have investigated the transport properties of an inertial Brownian particle moving in a one-dimensional, periodic and \emph{symmetric} potential which in addition is exposed to a harmonic ac driving as well as dichotomous noise of finite mean bias $\langle \eta(t) \rangle = f$ and compared them to the attributes of the same particle but subjected to a constant deterministic force $f$ instead of the random perturbation. 
We have presented the tailored parameter regime of the absolute negative mobility phenomenon such that when $f$ is replaced by $\eta(t)$ of equal average value, the transport properties of the driven Brownian particle are significantly improved, i.e. its negative velocity is enhanced, the velocity fluctuations are reduced and the Stokes efficiency becomes greater, each of them several times. Moreover, we studied the dependence of these quantifiers on  temperature of the system and 
revealed that this effect is assisted by  thermal fluctuations as it emerges only for specific interval of temperatures. Dichotomous noise, in contrast to other random perturbation which has been studied in this context before \cite{spiechPRE}, 
can assume both positive and negative values. Therefore we focused on the impact of this novel aspect of dynamics on the observed transport behaviour. We have demonstrated that by adjusting the parameters characterizing realizations of 
dichotomous noise it is possible to manipulate  the direction of transport occurring in this set-up. Furthermore, the most far-reaching conclusion of this work is that the 
phenomenon of non-equilibrium noise amplified efficiency of Brownian particles moving in periodic media may be still detected also for the case of bidirectional fluctuations. 
This mechanism may explain transport phenomena appearing in strongly fluctuating environments where instead of the deterministic forces random the perturbation without any unique direction operates.

Our results can be validated, for example, by use of a set-up consisting of the resistively and capacitively shunted Josephson junction working in experimentally accessible regimes. The Langevin equation (1) has a similar form  
as an equation of motion for the phase difference $\Psi=\Psi(t)$  between the macroscopic wave functions of the
Cooper pairs on both sides of the Josephson junction. The quasi-classical dynamics of the Josephson phase, which is well known in the literature as the Stewart-McCumber model \cite{kautz1996,stewart,mccumber} is described by the following equation
\begin{eqnarray} \label{JJ1}
\Big( \frac{\hbar}{2e} \Big)^2 C\:\ddot{\Psi} + \Big( \frac{\hbar}{2e} \Big)^2 \frac{1}{R} \dot{\Psi}
+ \frac{\hbar}{2e} I_0 \sin \Psi  = \frac{\hbar}{2e} I(t) + \frac{\hbar}{2e}
\sqrt{\frac{2 k_B T}{R}} \:\xi (t) .
\end{eqnarray}
The left hand side contains three additive current contributions: a displacement current due to the capacitance $C$ of the junction, a normal (Ohmic) current characterized by the resistance $R$ and a Cooper pair tunnel current characterized by the critical current $I_0$. In the right hand side $I(t)$ is an external current. Thermal fluctuations of the current are taken into account according to the fluctuation-dissipation theorem and satisfy the Johnson-Nyquist formula associated with the resistance $R$. There is an evident correspondence between two models: the coordinate $x=\Psi -\pi/2$, the mass $M=(\hbar /2e)^2 C$, the friction coefficient $\Gamma = (\hbar/2 e)^2(1/R)$ and the potential force $V'(x)$ translates to the Josephson supercurrent. The time-periodic force $G(t)$ in Eq. (1) corresponds to the external current $I(t)$. The velocity $v=\dot x$ translates to the voltage $V$ across the junction. Therefore all transport properties can be tested in the setup consisting of a resistively and capacitively shunted Josephson junction device.

\ack
This work was supported by the MNiSW program ”Diamond Grant” (J. S.) and NCN grant DEC-2013/09/B/ST3/01659 (J. {\L}.)

\end{document}